\documentclass[twocolumn,superscriptaddress]{revtex4}

\usepackage{amsfonts}
\usepackage{epsf}
\usepackage{graphics}

\begin{document}
\frenchspacing
\title{An ensemble approach to the analysis of weighted networks}
\author{S. E. Ahnert}
\affiliation{Institut Curie, CNRS UMR 144, 26 rue d'Ulm, 75248 Paris, France}
\author{D. Garlaschelli}
\affiliation{Dipartimento di Fisica, Universit\'a di Siena,
Via Roma 56, 53100 Siena, Italy}
\author{T. M. A. Fink}
\affiliation{Institut Curie, CNRS UMR 144, 26 rue d'Ulm, 75248 Paris, France}
\author{G. Caldarelli}
\affiliation{INFM-CNR Istituto dei Sistemi Complessi and Dipartimento di Fisica Universit\'a di Roma "La Sapienza" Piazzale Moro 2, 00185 Roma, Italy, and Centro Studi e Museo della Fisica Enrico Fermi, Compendio Viminale, 00185 Roma, Italy}

\begin{abstract}

\noindent
We present a new approach to the analysis of weighted networks, by providing a straightforward generalization of any network measure defined on unweighted networks, such as the average degree of the nearest neighbours, the clustering coefficient, the `betweenness', the distance between two nodes and the diameter of a network. All these measures are well established for unweighted networks but have hitherto proven difficult to define for weighted networks. Our approach is based on the translation of a weighted network into an ensemble of edges. Further to introducing this approach we demonstrate its advantages by applying the clustering coefficient constructed in this way to two real-world weighted networks.
\end{abstract}
\maketitle

\noindent
Weighted complex networks appear in many different contexts, for example when studying transport and traffic \cite{traffic, vespignani}, in the form of trade or communication networks, financial networks \cite{kertesz}, and collaboration networks \cite{scn}, to name a few. In addition, high-throughput technology has generated large amounts of biological data which can be interpreted in terms of weighted networks, such as networks of genetic regulation and transcription \cite{horvath} and protein interaction \cite{pin}. While such networks can now be generated relatively easy, the extraction of meaningful physical or biological information from these networks is a much more challenging task. For unweighted complex networks -- in which the entries of the adjacency matrix are restricted to zero and one -- a set of local and global measures on the network has been defined \cite{barabasi}, including the {\em degree} of a node, its {\em average nearest-neighbour degree} \cite{nearest} and its {\em clustering coefficient} \cite{clustering}. Further measures include the {\em distance} between two nodes, the related {\em diameter} of the network and the {\em betweenness} \cite{betweenness} of an edge or node. While the definition of such measures for unweighted networks is relatively straightforward, defining these measures for weighted networks is more difficult and has been the subject of recent research \cite{vespignani,onnela,horvath,newman,onnela2}. 

Here we introduce a new approach to this problem which allows for a straightforward generalization of any measure defined on an unweighted network to weighted networks. In addition we explicitly construct weighted versions of the clustering coefficient, the average degree of neighbours, the distance between two nodes and the diameter of the network. We compare this newly constructed clustering coefficient to a weighted clustering coefficient in the literature and to a version used in unweighted networks. The data sets we use for this comparison are aviation passenger data within the EU, which constitutes an almost fully connected network, and the network formed by neighbouring letters in the English language.

{\em Ensemble networks}---%
The basis of our approach is to find a continuous bijective map $M : \mathbb{R} \rightarrow [0,1]$ from the real numbers to the interval between 0 and 1, which maps the weights $w_{ij} \in \mathbb{R}$ to a quantity $p_{ij} \in [0,1]$. A simple example of such a map is a linear normalization of the weights:
\begin{equation}\label{norm}
p_{ij} = {w_{ij} - {\rm min} (w_{ij}) \over {\rm max}(w_{ij}) - {\rm min}(w_{ij})}
\end{equation}
This simple normalization maps ${\rm min} w_{ij}$ to zero. This is often acceptable in the case of a distance matrix, but if there are many edges with weight ${\rm min} w_{ij}$, one should introduce a parameter $\epsilon \ll 1$, such that:
\begin{equation}\label{norm}
p_{ij} = {w_{ij} - {\rm min} (w_{ij}) + \epsilon \over {\rm max}(w_{ij}) - {\rm min}(w_{ij}) + \epsilon}
\end{equation}
Many other more sophisticated maps are imaginable and the final choice of map depends on the properties of the physical system underlying the network and the resulting distribution of weights. Appropriately chosen maps can deal with all variants of weighted networks including those with negative weights, and with differing interpretations of $w_{ij} = 0$ as meaning 'no edge' or as a physical weight. We will return to the topic of map choice below. 

The ideas we introduce in this paper are based on an interpretation of the matrix ${\bf P}$ with entries $\{p_{ij}\}$ as a matrix of {\em probabilities}. These probabilities can be interpreted as an {\em ensemble of edges}, or more concisely, an {\em ensemble network}. Thus, just as any binary square matrix can be understood as an unweighted network and any real square matrix corresponds to a weighted network, any square matrix with entries between 0 and 1 corresponds to an ensemble network. If we sample each edge of the ensemble network exactly once, we obtain an unweighted network which we term a {\em realization} of the ensemble network. In particular, $p_{ij}$ is the probability that the edge between nodes $i$ and $j$ exists.
These concepts are valid both for directed networks, with any $p_{ij} \in [0,1]$, and undirected networks, for which $p_{ij} = p_{ji}$, so that the matrix is symmetric. Note that, while some specific weighted networks discussed in the literature have probabilities as their weights \cite{horvath,grindrod}, a general framework for the analysis of weighted networks, based on the transformation of weights to probabilities, has to our knowledge not been proposed. In a real-world weighted network, the original weights can represent almost any physical quantity, such as the strength of a collaboration between two scientists, or the number of passengers traveling between two countries. This is why we use a map $M$ to translate the original weights into probabilities. Doing so does not destroy any of the topological information contained in the weights and connections, but allows us to analyze this information in the unifying framework which the probabilities provide. Furthermore, in many cases of real-world weighted networks, the transformation of weights to probabilities has a physical meaning. Examples include flow networks of traffic and transport, communications networks as well as collaboration networks. In all these, the interactions between nodes involve the transfer of a discrete unit (e.g. passengers, currency or data packets) over a given period of time. Thus the weight, representing the number of units transferred, is directly related to the probability of observing the transfer of a unit at a given point in time. 

In the framework of ensemble networks any existing measure on unweighted networks can be turned into an equivalent measure on weighted networks. A suitable choice of map $M$ depends on the distribution of weights. For example, in both real-world networks which we analyze in this paper the original weights $w_{ij}$ take values across several orders of magnitude, so that we chose the $p_{ij}$ to be the normalized logarithms of the original weights, rather than the normalized weights themselves.  

{\em Polynomials of adjacency matrix entries}---%
All measures on unweighted networks can be written as functions of the entries $a_{ij}$ of an adjacency matrix ${\bf A}$. In fact, generally they can be written as a polynomial of these entries, or a simple ratio of such polynomials. Note that, for an unweighted network, $a_{ij} = a_{ij}^m$ for all positive integers $m > 0$, so that these polynomials are of first order only. Consider a general first-order polynomial, which can be written fully expanded as:
\[
f({\bf A}) = \sum_{q = 0}^{2^{N^2}} C_q \prod_{j,k = 0}^{N} a_{jk}^{b(q)_{jk}} 
\]
where $N$ is the number of nodes, the $C_q$ are real coefficients and the $b(q)_{jk}$ are a set of boolean matrices specifying which adjacency matrix entries appear in each term of the polynomial.  
The probability $P_q$ that $\prod_{j,k = 0}^{N} a_{jk}^{b(q)_{jk}} = 1$ in a given realization ${\bf A}$ is simply $P_q = \prod_{j,k = 0}^{N} p_{jk}^{b(q)_{jk}}$. Thus, due to the linearity of the polynomial, the average $\bar{f}({\bf P})$ of $f$ over the ensemble network realizations is:
\begin{equation}\label{poly}
\bar{f}({\bf P}) = \sum_{q = 0}^{2^{N^2}} C_q \prod_{j,k = 0}^{N} p_{jk}^{b(q)_{jk}} = f({\bf P})
\end{equation}
This means that the value of a polynomial function $f$ of the entries of an unweighted network ${\bf A}$, averaged over the realizations of a given ensemble network ${\bf P}$ is equal to the value of the polynomial of the ensemble network adjacency matrix itself. We will illustrate the power of this result in the following sections.

{\em Constructing the measures}---%
Our approach allows for the construction of weighted network measures from their unweighted counterparts. As almost all existing unweighted measures are for undirected networks, the measures we construct in the remainder of this Letter are also undirected. In general however our method is equally well suited to the transformation of any measure for directed, unweighted networks into one for directed and weighted networks. 
The degree $k_i$ of a given node $i$ in an unweighted network with adjacency matrix elements $a_{ij}$ is the number of its neighbours, and is written as $k_i = \sum_{j} a_{ij}$. 
In a weighted network with elements $w_{ij}$ the corresponding quantity has been termed the {\em strength} of the node $i$, denoted as $s_i$, which consists of the sum of the weights: $s_i = \sum_{j} w_{ij}$.
In an ensemble network, the corresponding sum over the edges attached to a particular node gives the {\em average degree} of node $i$ {\em across realizations}, denoted as $\bar{k_i}$ and given by $\bar{k_i} = \sum_{j} p_{ij}$.
 
It is important to note that while the strength of a node in a weighted network may have meaning in the context of the network, $\bar{k_i}$ has a universal meaning, regardless of the original meaning of the weights. 
Now consider the total number of edges $n$ in a network -- also referred to as its {\em size} -- given by $n = \sum_{ij} a_{ij}$ in the directed case and half this value in the undirected case where $a_{ij} = a_{ji}$.
Replacing $a_{ij}$ by $p_{ij}$ again gives us the average size $\bar{n}$ of the realizations of the ensemble network, which is simply $\bar{n} = \sum_{ij} p_{ij}$ (or half this value for the undirected case).

A more complex measure in unweighted networks is the average degree of the nearest neighbours $k^{nn}_i$, which is the number of neighbours of $i$'s neighbours, divided by the number of neighbours of $i$ \cite{nearest}:
\[
k^{nn}_i = {\sum_j k_j \over k_i} = {\sum_{j,k} a_{ij} a_{jk} \over \sum_{j} a_{ij}} 
\]
where $j \neq i$ in the sums. By rewriting $k^{nn}_i$ solely in terms of the $a_{ij}$, this generalizes to ensemble networks in a very straightforward manner:
\[
k^{nn,e}_i = {\sum_{j,k} p_{ij} p_{jk} \over \sum_{j} p_{ij}} 
\]
This measure $k^{nn,e}$ is simply a ratio of averages: the average number of neighbours of $i$'s neighbours over the average number of $i$'s neighbours. 

For unweighted networks the {\em clustering coefficient} of a node $i$ has been defined \cite{clustering} as:
\begin{equation}\label{uclusteringeq}
c_i = {\sum_{j,k} a_{ij} a_{jk} a_{ik} \over k(k-1)/2} = {\sum_{j,k} a_{ij} a_{jk} a_{ik} \over \sum_{j,k} a_{ij} a_{ik}} 
\end{equation}
where $k \neq j \neq i \neq k$ in the sums. This corresponds to the number of triangles in the network which include node $i$, divided by the number of pairs of bonds including $i$, which represent potential triangles. Using the ensemble approach with its normalized weights this generalizes straightforwardly to:
\begin{equation}\label{clusteringeq}
c^e_i = {\sum_{j,k} p_{ij} p_{jk} p_{ik} \over \sum_{j,k} p_{ij} p_{ik}} 
\end{equation}
which can be read as the average number of triangles divided by the average number of bond pairs. In modified form, this clustering coefficient has appeared in the very recent literature \cite{horvath} but without connection to a general approach to the construction of weighted network measures based on a general mapping from weights to probabilities. 
Note that $k^{nn,e}$ and $c^e_i$ are {\em not} the averages of $k^{nn}_i$ and $c_i$ over the ensemble. We will address this subtlety below.

As an example of the power of eq. (\ref{poly}), consider the {\em distance} $d_{ij}$ (i.e. the shortest path) between two nodes $i$ and $j$ in an unweighted $N$-node network, represented entirely as a function of adjacency matrix entries:
\[
d_{ij} ({\bf A})= a_{ij} + (1 - a_{ij}) \sum_{m = 1}^N (m+1) \, \alpha^{(m)}_{ij} ({\bf A}) \, \beta^{(m)}_{ij} ({\bf A}) 
\]
where $\alpha^{(m)}_{ij} ({\bf A}) = \prod_{q = 1}^{m-1} [1 - \beta^{(q)}_{ij}]$, with:
\[
\beta^{(m)}_{ij} ({\bf A}) = 1 - \prod_{k_1,...,k_m} \left(1 - a_{ik_1} a_{k_mj} \prod_{l = 1}^{m-1} a_{k_lk_{l+1}}\right)
\]
where all $\prod$ without a range are equal to one. 
As $d_{ij}$ is a first-order polynomial in $a_{ij}$ -- the elements of the adjacency matrix ${\bf A}$ -- we know immediately from eq. (\ref{poly}) that the average distance in the ensemble network will be $\bar{d}_{ij} ({\bf P}) = d_{ij}({\bf P})$. Thus we have defined a distance measure on weighted networks without having to define a pairwise distance function of the edge weights (such as, for example, $d_{ij} = (w_{ij})^{-1}$ \cite{scn}).

Similarly, the {\em diameter} of an unweighted network, defined as the maximum distance $D({\bf A}) = {\rm max} \, d_{ij} ({\bf A})$ between two nodes out of all pairs of nodes $i,j$ can be written as a first-order polynomial:
\[
D({\bf A}) = \prod_{p,q} a_{pq} + \sum_{m = 1}^N (m+1) \, \zeta^{(m)} ({\bf A}) \, \xi^{(m)} ({\bf A})  
\]
where $\zeta^{(m)} ({\bf A}) = \prod_{q = 1}^{m-1} [1 - \xi^{(q)} ({\bf A})]$ 
and $\xi^{(m)} ({\bf A}) = \prod_{i,j} \beta^{(m)}_{ij} ({\bf A})$.
This expression allows us to straightforwardly calculate the average diameter $\bar{D}({\bf P}) = D({\bf P})$ of the ensemble network. 

\begin{figure}
\begin{tabular}{c}
\hspace{-0.6cm}\scalebox{0.3}[0.3]{\rotatebox{-90}{\epsfbox{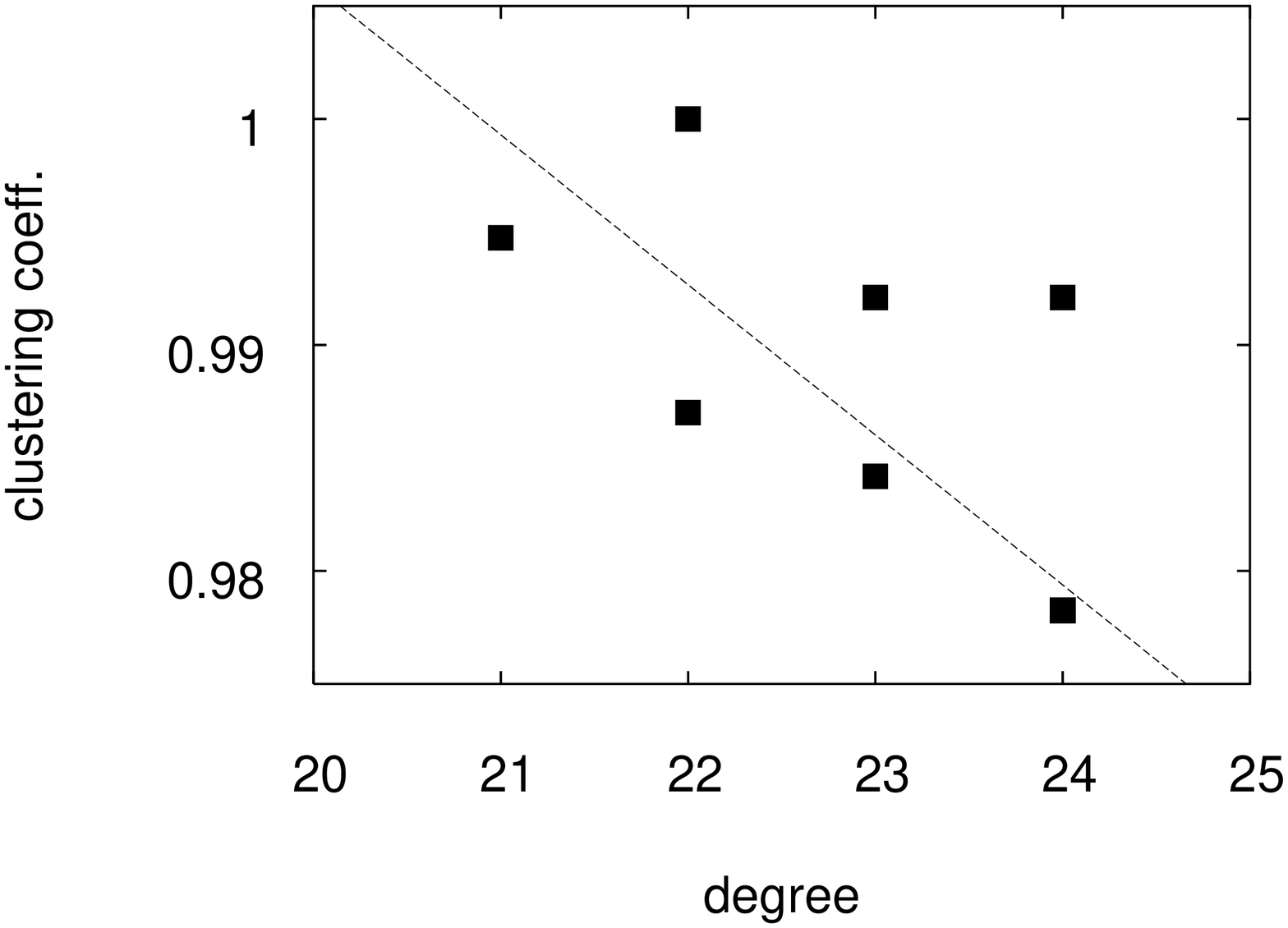}}}\cr
\hspace{-0.8cm}\scalebox{0.32}[0.3]{\rotatebox{-90}{\epsfbox{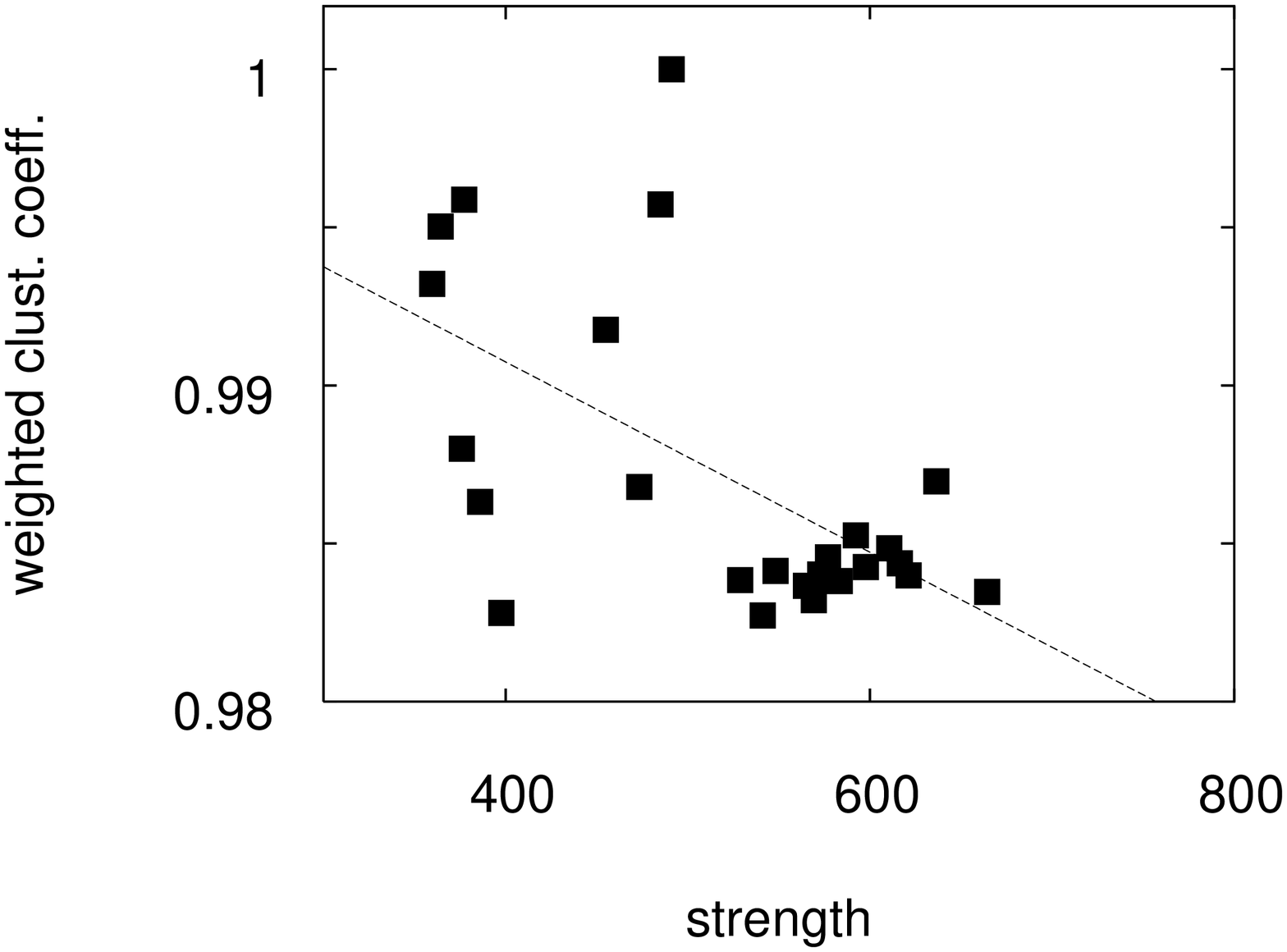}}}\cr
\hspace{-0.3cm}\scalebox{0.3}[0.3]{\rotatebox{-90}{\epsfbox{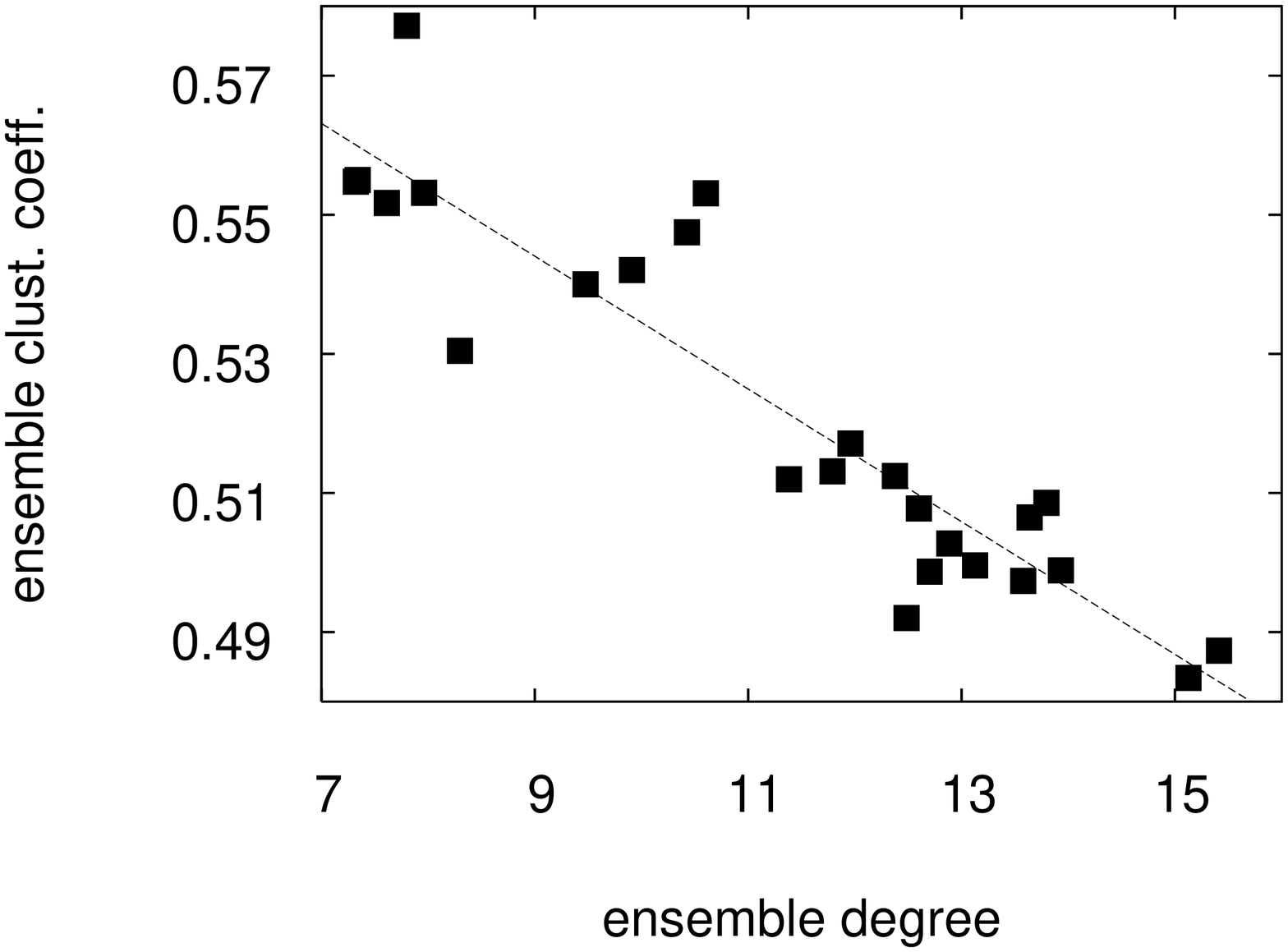}}}\cr
\end{tabular}
\caption{Analysis of the network of air travel passengers within the 25 member states of the EU. This network is almost fully connected. TOP: Unweighted clustering coefficient versus degree.All 25 data points are projected onto 7 locations, as a result of the information loss due to discarding the weights, and because the network is almost fully connected. MIDDLE: Clustering coefficient as proposed in the literature \cite{vespignani} versus strength. This ``mixed'' clustering coefficient is a function of unweighted and weighted quantities. No clear relationship is evident, again because the network is almost fully connected. BOTTOM: Ensemble clustering coefficient versus ensemble degree. Unlike the other two approaches, those derived using the ensemble quantities exhibit a clear negative linear relationship. The lines are lines of best fit. Note that the absolute scale of the ensemble clustering coefficient $c^e_i$ depends on the choice of the map $M$ from weights to probabilities, which makes the relative values of $c^e_i$ more important than the absolute ones.}\label{eu}
\end{figure} 

\begin{figure}
\begin{tabular}{c}
\hspace{-0.6cm}\scalebox{0.3}[0.3]{\rotatebox{-90}{\epsfbox{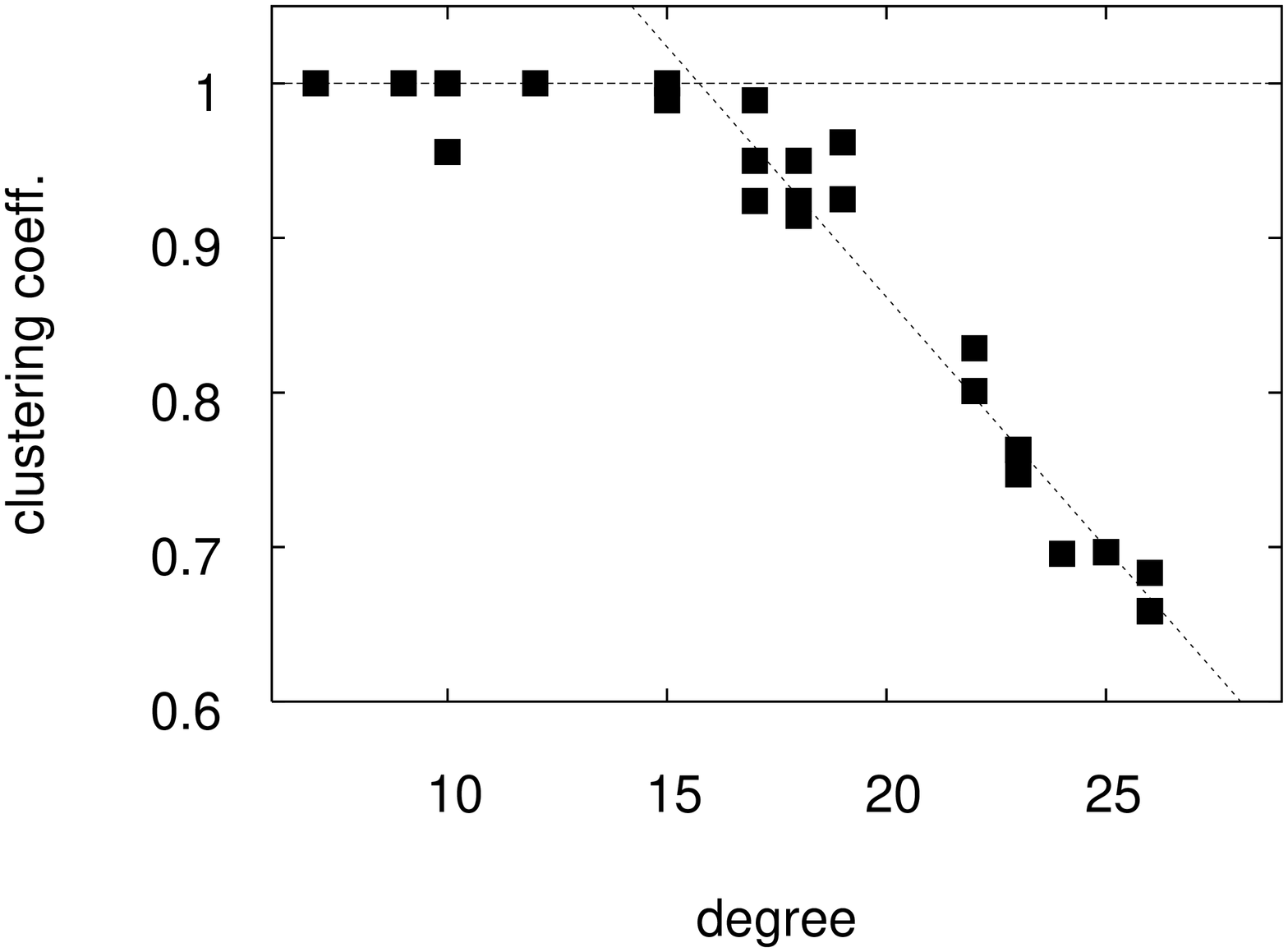}}}\cr
\hspace{-0.3cm}\scalebox{0.3}[0.3]{\rotatebox{-90}{\epsfbox{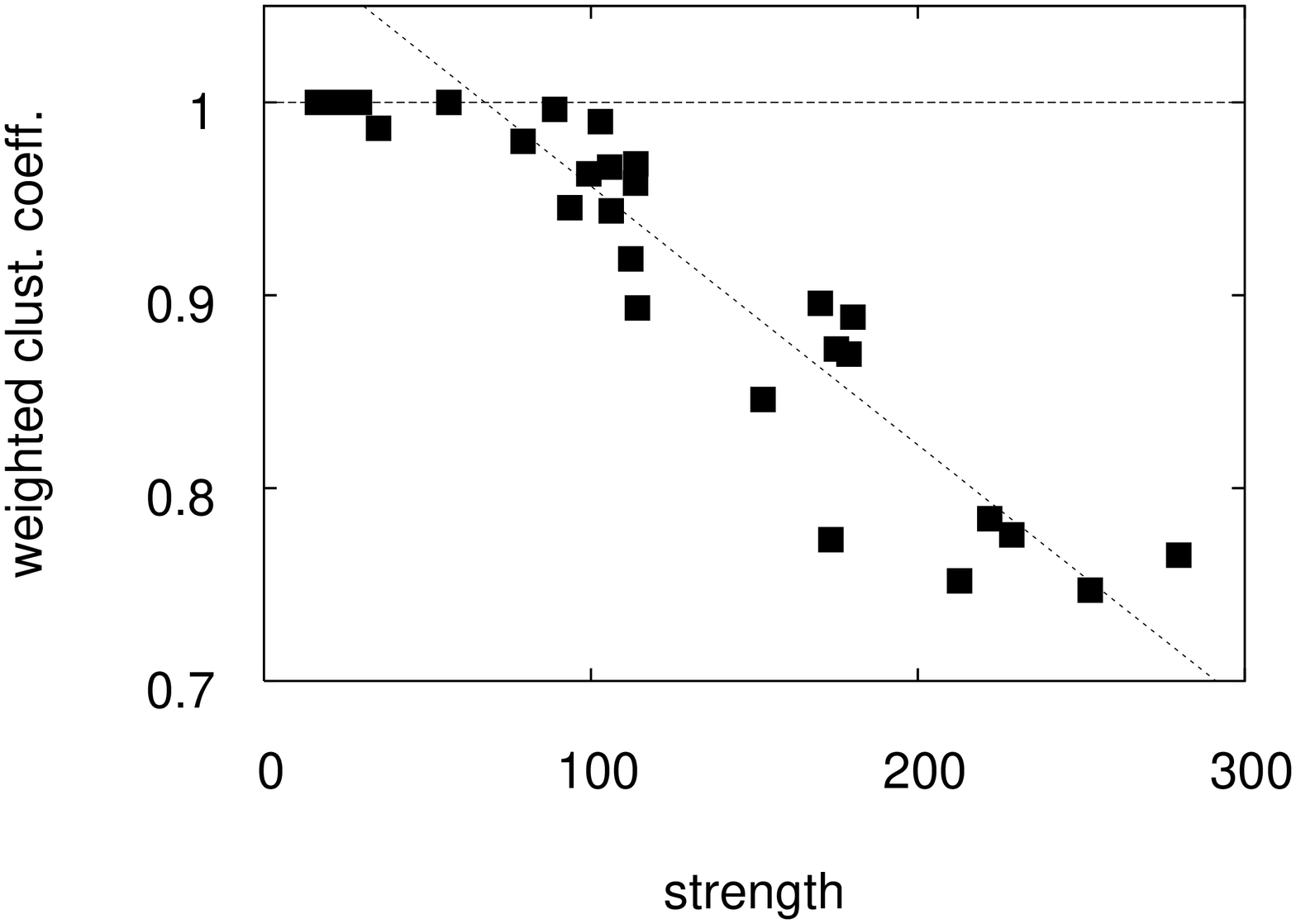}}}\cr
\hspace{-0.3cm}\scalebox{0.3}[0.3]{\rotatebox{-90}{\epsfbox{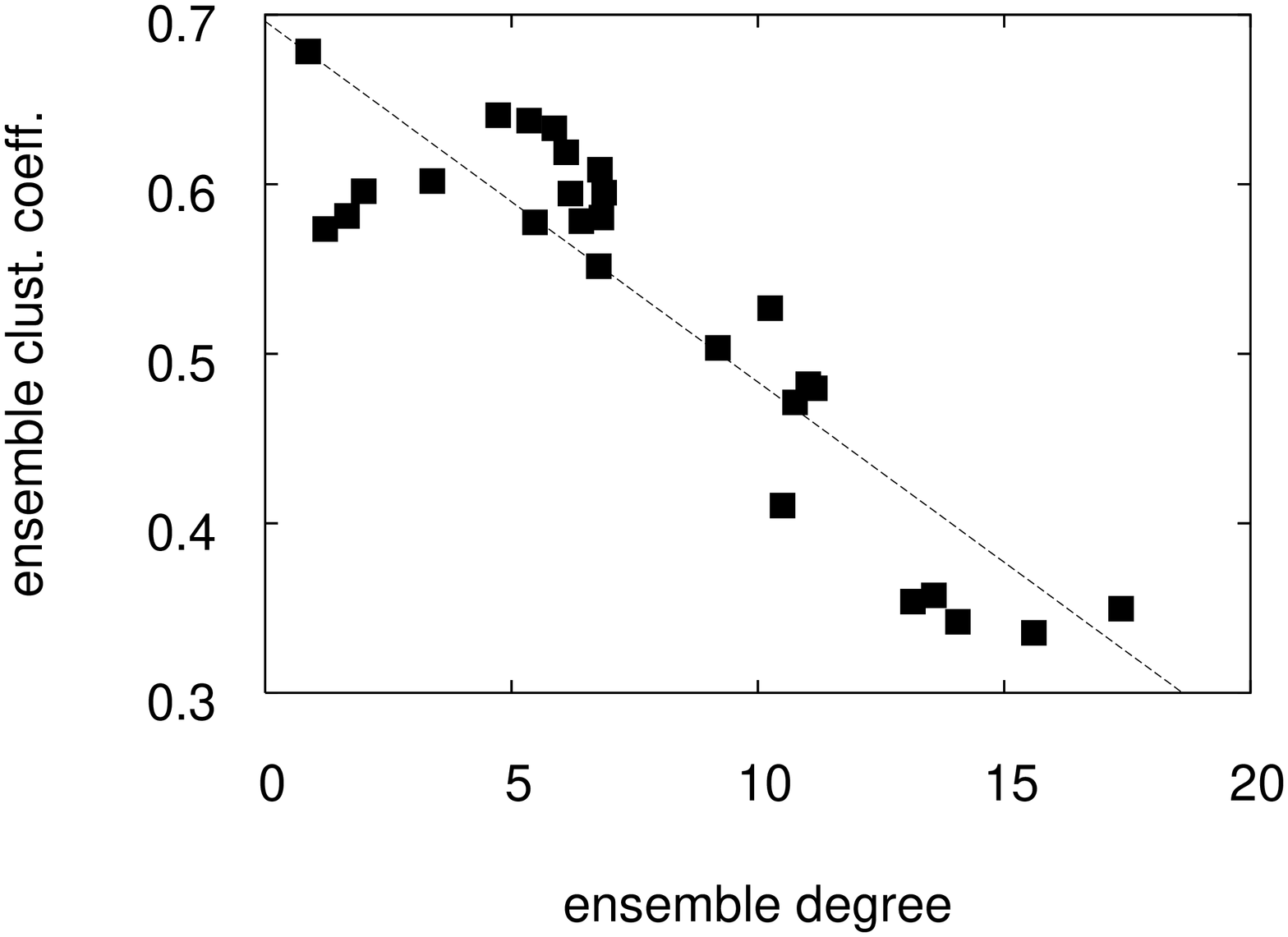}}}\cr
\end{tabular}
\caption{Analysis of the weighted network formed between the 26 letters of the alphabet and the space between words \cite{lettersref}. TOP: Unweighted clustering coefficient versus degree. MIDDLE: Clustering coefficient as proposed in the literature \cite{vespignani} versus strength. This ``mixed'' clustering coefficient is a function of unweighted and weighted quantities. BOTTOM: Ensemble clustering coefficient versus ensemble degree. The ensemble approach makes use of all information contained in the weights, while the two others lose some of the information, as is shown by the plateau which both exhibit on the left side of the plots. The diagonal lines are lines of best fit for data points below the plateau. Note that the absolute scale of the ensemble clustering coefficient $c^e_i$ depends on the choice of the map $M$ from weights to probabilities, which makes the relative values of $c^e_i$ more important than the absolute ones.}\label{letters}
\end{figure}

Another measure, the {\em betweenness} \cite{betweenness} of a node $i$ or an edge $(i,j)$, is the number of different shortest paths in the network which run through $i$ or $(i,j)$s in the network. Like measures such as the distance and diameter, the betweenness can also be generalized to the weighted case by simply replacing the $a_{ij}$ by $p_{ij}$. As the expressions in terms of adjacency matrix entries are rather involved, we do not give them here explicitly. 

Some measures on unweighted networks, such as the average neighbour degree $k_{nn}$ and the clustering coefficient $c_i$ are {\em ratios} of two adjacency matrix polynomials $f$ and $g$, which in general can be written as $h({\bf A}) = {f({\bf A})/g({\bf A})}$. Now we can {\em define} the quantity $h^e({\bf P}) \equiv {\bar{f}({\bf P}) / \bar{g}({\bf P})} = h({\bf P})$.
But, as was pointed out above, this quantity is no longer an average of $h({\bf A})$ itself (which would be denoted $\bar{h}({\bf P})$). This gives us two distinct classes of measures: The first contains measures which can be written in polynomial form, and for which the ensemble version gives the average across realizations. These measures represent countable, integer quantities of the network, such as the number of neighbours, the number of triangles, the length of the shortest path between $i$ and $j$, and so on. The second class are measures which are ratios of polynomials, such as the average nearest-neighbour degree or the clustering coefficient. The ensemble network version of these measures gives the ratio of the averages. 

All measures constructed with the ensemble approach are only functions of the normalized weights $p_{ij}$, not of the elements of an unweighted adjacency matrix $a_{ij}$ or of the degree $k$. This distinguishes the ensemble measures from measures proposed for weighted networks in the literature, such as the weighted clustering coefficient $c_i^w$:
\begin{equation}\label{vesp}
c_i^w = {1 \over s_i (k_i - 1)} \sum_{j,k} {(w_{ij} + w_{ik}) \over 2} a_{ij} a_{ik} a_{jk}
\end{equation}
and the weighted average nearest-neighbour degree $k_{nn,i}^w$:
\begin{equation}\label{vesp2}
k_{nn,i}^w = {1 \over s_i} \sum_{j=1}^N a_{ij} w_{ij} k_j
\end{equation}

Both are defined in \cite{vespignani}. Due to their construction, these measures cannot be used for the analysis of fully connected weighted networks, as $k_{nn,i}^w = 1$ and $c_i^w = 1$ for all nodes $i$ in such networks. Fully connected weighted networks form an important class of complex networks, for example in the form of the (virtually fully-connected) EU air travel network which we analyze in this letter. Furthermore {\em any} matrix of similarities or distances between a number of objects - such as for instance microarray data series in biological experiments - can be treated as a fully connected weighted network, and thus can be analyzed using the ensemble approach, but not with approaches such as eq. (\ref{vesp}) and (\ref{vesp2}), which are ``mixed'' in the sense that they make use of both the unweighted and weighted adjacency matrix entries. 

{\em Analyzing real-world weighted networks}---%
In the following we demonstrate some of the advantages which the ensemble approach has over unweighted network measures, as well as over mixed weighted network measures. We do this by applying the ensemble clustering coefficient of eq. (\ref{clusteringeq}) to two real-world networks. The first is the network of passengers travelling by air within the EU during 2004 \cite{eurostat}. The second is a network of letters in the English language, where the weight of the edge between two letters is determined by the freqency at which they appear next to each other in the English language \cite{lettersref}. Both networks include edges which lead from a node to itself. The network of letters has 485 edges between 27 nodes (the alphabet and space), and therefore is 62.8\% connected, while the EU network with 607 connections between the 25 member states of the EU is almost fully (97.1\%) connected. 

In Fig. \ref{eu} we show the analysis of the EU air travel network using three different clustering coefficients: the unweighted clustering coefficient $c_i$ of eq. (\ref{uclusteringeq}) \cite{clustering}, the mixed weighted clustering coefficient $c_i^w$ of eq. (\ref{vesp}) with weighted and unweighted components from the literature \cite{vespignani} and the clustering coefficient $c^e_i$ of eq. (\ref{clusteringeq}) derived from the ensemble approach. These quantities are plotted against the degree $k$ (in the unweighted case), the strength $s_i$ for the mixed approach, and the ensemble degree $\bar{k}_i$ in for the ensemble approach. As this network is almost fully (97.1\%) connected, the difficulty of the unweighted and mixed approaches becomes apparent: For the unweighted case, the 25 nodes of the network are mapped to just 7 points, representing the information lost by dropping the weights. In the mixed case, little can be deduced about the relationship between the clustering coefficient $c_i^w$ and the strength $s_i$. The ensemble approach on the other hand reveals a clear negative linear relationship between the ensemble clustering coefficient $c^e_i$ of eq. (\ref{clusteringeq}) and the ensemble degree $\bar{k}_i$. Note that the absolute values of the ensemble clustering coefficient do not mean very much, as they are dependent on the map $M$. It is their relative values which carry the information, and these are largely independent of the choice of map $M$, as long as it is bijective. Countries with a large number of air passengers travelling in and out have a high ensemble degree $\bar{k}_i$ but also a low ensemble clustering coefficient $c^e_i$, as the many countries they are connected to strongly are mostly not well-connected themselves. Thus these nodes with low $c^e_i$ are surrounded by few triangles in any given ensemble realization, but many potential triangles in the form of pairs of edges. The inverse argument is true for nodes with a low ensemble degree $\bar{k}_i$, as any two neighbours of such a node are more likely to be strongly connected. For example, the two countries at the bottom right of the plot (high $\bar{k}_i$, low $c^e_i$) are the UK and Germany, while the top left corner (low $\bar{k}_i$, high $c^e_i$) contains Lithuania, Estonia and Slovakia.

In Fig. \ref{letters} we show the analysis of the letter network using the same three clustering coefficients. As the letter network is less than two-thirds (62.8\%) connected, the unweighted and mixed approaches do not encounter the difficulties associated with fully connected networks. However, if there are clusters in the network which are fully connected on a local scale -- such that all neighbours of a given node are fully connected -- these approaches again cannot differentiate any further between such nodes. In both the unweighted and mixed cases the letters Q, Z, J and V are affected, as these letters only have few neighbours, which are fully connected among themselves, making the unweighted and mixed clustering coefficients equal to one. In Fig. \ref{letters} these four letters are represented by the four data points on the plateau which appears in the plots for the unweighted and mixed measures. No information however is lost with the ensemble approach, which again shows a clear negative linear relationship between ensemble clustering coefficient and ensemble degree. As before, the implication of this is that nodes with many strong connections -- in this case the vowels A, E, I, O and U, which are located at the bottom right of the plots in Fig. \ref{letters} -- have neighbours which are weakly connected among each other. These are the consonants, which are mostly located in the top left corner (low $\bar{k}_i$, high $c^e_i$). 

{\em Conclusion}---%
We have introduced a general approach for the construction of measures on weighted networks, by introducing the concept of an ensemble network, in which every edge has a probability $p_{ij}$ of existing. By transforming a weighted network into an ensemble network, any of the numerous measures which have been defined for unweighted networks can be straightforwardly generalized to weighted networks. Using the clustering coefficient constructed in this way as an example we demonstrate that these measures on weighted networks can reveal the additional topological information given in the weights, in particular for fully connected networks.

\end{document}